\documentclass[conference,letterpaper]{IEEEtran}
\IEEEoverridecommandlockouts
\usepackage{booktabs}
\usepackage{algorithm}
\usepackage[noend]{algpseudocode} 
\usepackage[hidelinks]{hyperref}
\usepackage[letterpaper,
            left=1.05in,   
            right=1.05in,  
            top=0.75in,    
            bottom=1.05in  
           ]{geometry}

\usepackage{amsmath,amssymb,amsfonts}
\usepackage{algpseudocode}
\usepackage{graphicx}
\def\BibTeX{{\rm B\kern-.05em{\sc i\kern-.025em b}\kern-.08em
    T\kern-.1667em\lower.7ex\hbox{E}\kern-.125emX}}

\makeatother

\begin{document}

\title{gNB-based Local Breakout for URLLC in industrial 5G}

\author{
\IEEEauthorblockN{Rajendra Paudyal\textsuperscript{*}, Rajendra Upadhyay\textsuperscript{\dag}, Al Nahian Bin Emran\textsuperscript{\ddag}, Duminda Wijesekera\textsuperscript{\dag}}
\IEEEauthorblockA{
\text{Department of Computer Science\textsuperscript{*}}, 
\text{Department of Cyber Security\textsuperscript{\dag}}, 
\text{Department of Information Technology\textsuperscript{\ddag}}  \\
\text{George Mason University} \\
Fairfax, VA, USA \\
\text{\textsuperscript{*}rpaudyal@gmu.edu, \textsuperscript{\dag}rupadhya@gmu.edu, \textsuperscript{\ddag}abinemra@gmu.edu, \textsuperscript{\dag}dwijesek@gmu.edu}
}
}



\maketitle

\begin{abstract}
Industrial URLLC workloads-coordinated robotics, automated guided vehicles, machine-vision collaboration require sub-5\,ms latency and five-nines reliability. In standardized 5G Multicast/Broadcast Services, intra-cell group traffic remains anchored in the core using MB-SMF/MB-UPF, and the Application Function. This incurs a core network path and packet delay that is avoidable when data transmitters and receivers share a cell. We propose a gNB-local multicast breakout that pivots eligible uplink flows to a downlink point-to-multipoint bearer within the gNB, while maintaining authorization, membership, and policy in the 5G core. The design specifies an eligibility policy, configured-grant uplink. 3GPP security and compliance are preserved via unchanged control-plane anchors. A latency budget and simulation indicate that removing the backhaul/UPF/AF segment reduces end-to-end latency from \(\approx\)6.5-11.5\,ms (anchored to the core) to \(\approx\)1.5-4.0\,ms (local breakout), producing sub-2\,ms averages and a stable gap \(\approx\)10\,ms between group sizes. The approach offers a practical, standards-aligned path to deterministic intra-cell group dissemination in private 5G. We outline multi-cell and prototype validation as future work.
\end{abstract}


\section{Introduction}
\label{sec:introduction}

Industry~4.0 and emerging Industry~5.0 settings blend cyber-physical systems (CPS), AI, and digital twins to achieve adaptive, resilient, and efficient manufacturing. Examples include: (i) coordinated multi-robot assembly with sub-frame motion updates; (ii) UAV swarms requiring fast group state dissemination; (iii) machine vision alarms that must reach multiple controllers within a control cycle and (iv) human robot collaboration where safety relevant signals require deterministic delivery~\cite{3gpp.22.104,ericsson.i40}. Private 5G deployments are attractive due to licensed-grade interference resilience, mobility, and QoS controls. However, for \emph{intra-cell} group communication, the standard 5G Multicast/Broadcast Service (MBS) model routes multicast through MB-UPF in the core~\cite{3gpp.23.247}. Even on a campus with short fiber backhaul, traversing gNB  $\rightarrow$ UPF / MB-UPF $\rightarrow$ AF / application $\rightarrow$ gNB adds queueing and processing that can undermine URLLC targets~\cite{3gpp.23.501,3gpp.23.502}.

We address this gap with \emph{RAN-local multicast breakout}. The gNB forwards specific uplink flows from group members to a downlink multicast bearer without sending payloads to the core. Only the user-plane packets are rerouted towards the downlink multicast channel, where all other control-packets are handled by the core. This retains core-driven authorization and session control. Only control-plane packets are routed towards the core by the gNB; data plane qualified packets qualify for local breakout are rerouted by the gNB to the multicast group. This reduces latency to single-digit milliseconds for intra-cell groups. The concept of local breakout is taken from the paper~\cite{raj}. 


\section{Background}
\subsection{MBS in 5G}
In the Third Generation Partnership Project (3GPP) Release 17, MBS are introduced as a key enhancement to the 5G system architecture. The release established efficient PTM communication for applications such as public safety, vehicle-to-everything (V2X), Internet Protocol Television (IPTV), and group communications. This architecture builds on the existing 5G framework defined in Technical Specification (TS) 23.501 by incorporating new network functions to support both multicast and broadcast modes. It emphasizes resource efficiency and compatibility with non-roaming scenarios~\cite{acm-2024-chukhno,3gpp.23.247,garro-2025}. Specifically, the Multicast/Broadcast Session Management Function (MB-SMF) is responsible for session control, including the creation, activation, deactivation, modification and deletion of MBS sessions, as well as the allocation of Temporary Mobile Group Identities (TMGIs), the derivation of Quality of Service (QoS) parameters, and coordination with the Access and Mobility Management Function (AMF) for the allocation of resources from the Radio Access Network (RAN). Complementing this, the Multicast/Broadcast user-plane Function (UPF) serves as the MBS session anchor. It handles user-plane replication and distribution by receiving a single copy of MBS data from the Application Function (AF) or Multicast/Broadcast Service Transport Function (MBSTF) and forwarding it using General Packet Radio Service Tunneling Protocol User Plane (GTP-U) tunnels to Next Generation RAN (NG-RAN) nodes or other UPFs for downstream delivery. 

RAN plays a key role in optimizing delivery on the radio interface, dynamically selecting between PTM and point-to-point (PTP) modes per User Equipment (UE) based on factors such as UE density, radio conditions, and mobility~\cite{Shrivastava2025}. In PTM mode, the NG-RAN transmits a single copy of MBS data over a Multicast Radio Bearer (MRB) to multiple UEs, leveraging multicast transport for shared delivery. This enhances spectral efficiency in scenarios with high UE concentrations~\cite{LANDROVE2024,3gpp.23.247}. In contrast, the PTP mode involves separate transmissions to individual UEs via unicast Protocol Data Unit (PDU) sessions, ensuring reliability in sparse or edge coverage areas, with seamless switching between modes to maintain service continuity during handovers~\cite{Shrivastava2025}. This architecture demonstrates robustness and scalability across multiple cells, supporting features like location-dependent services, multiple QoS flows (Guaranteed Bit Rate and non-Guaranteed Bit Rate), and mobility handling through transitions between 5G Core (5GC) shared and individual delivery methods. 

However, a notable limitation lies in the centralization of user-plane replication within the core network at the MB-UPF, which acts as a centralized anchor for data ingress and distribution, even when UEs are co-located within a single cell~\cite{3gpp.23.247}. While this design reuses existing 5G entities to minimize deployment costs and supports efficient multicast transport where available, it can introduce inefficiencies in localized scenarios, as replication for individual delivery or unicast tunnels occurs in the core rather than being distributed closer to the RAN, potentially increasing latency and backhaul load~\cite{Shrivastava2025}.

\begin{figure}[t]
  \centering
  \includegraphics[width=0.92\linewidth]{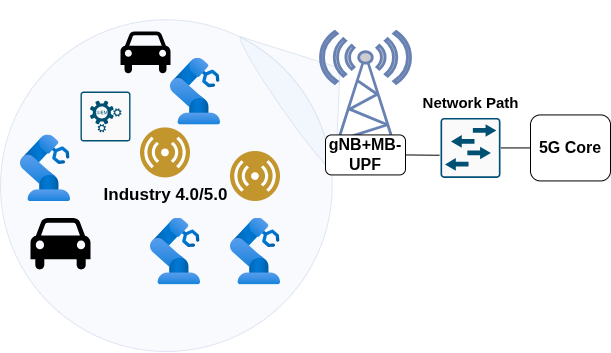}
  \caption{Architectural overview of local multicast}
  \label{fig:arch1}
\end{figure}

\subsection{URLLC for Industrial Control}
\label{ssec:URLLCforIndustry}

In the context of 5G and beyond networks, 3GPP has established stringent service requirements for cyber-physical control applications. Particularly in vertical domains such as industrial automation, where URLLC is paramount to ensure deterministic performance with end-to-end latencies as low as 5 ms and reliabilities exceeding 99.9999\% for packet transmissions~\cite{3gpp.22.104}. These requirements, outlined in TS 22.104, address scenarios involving real-time feedback loops in machine-to-machine interactions. It emphasizes the need for robust communication services to support closed-loop control systems without compromising safety or efficiency. The foundational building blocks of URLLC in 5G are shortened transmission time intervals (TTIs) by reduced slot durations, configured-grant enabling grant-free uplink access to minimize scheduling delays, packet duplication across multiple paths for enhanced reliability, robust modulation and coding schemes (MCS) to combat channel impairments, and mini slots for flexible sub slot transmissions~\cite{coll-2022}. However, extending these mechanisms to group dissemination scenarios, such as multicast URLLC (mURLLC) for synchronized control of device clusters, introduces additional complexities in feedback handling and scheduling, as dynamic resource allocation must balance individual user conditions with group wide efficiency, potentially increasing computational overhead and risking latency violations in dense environments.

Figure \ref{fig:arch1} illustrates an Industry 4.0/5.0 application connected to the gNB with local user-plane function “gNB+MB-UPF”. Industrial endpoints inside the cell (robots, AGVs, controllers, vehicles) generate group updates that, when policy and membership allow, are pivoted at the gNB directly into a downlink point-to-multipoint (PTM) bearer and transmitted once to all in-cell receivers bypassing the network path to the 5G Core. While control-plane functions remain anchored in the core. This avoids the gNB → UPF/AF → gNB path, shortening delivery to UE → gNB → group UEs and driving the latency gains.

\subsection{Local Breakout}
\label{ssec:localBreakout}

Local breakout in cellular networks is a pivotal strategy to optimize traffic routing by anchoring data flows at the network edge. It minimizes backhaul traversal and reduces end-to-end latency. In private 5G deployments, Local breakout enables localized processing by integrating media and control-plane functions directly at the gNB as defined in 3GPP TS 23.501. This approach is particularly advantageous for latency-sensitive applications such as industrial IoT. The proximity to the data source improves performance while alleviating core network congestion. This pivoting traffic flow path reduces backhaul load and supports improving scalability. Furthermore, 3GPP TS 33.501 ensures that security and policy enforcement, including authentication, authorization, and QoS management, are preserved at the core through standardized mechanisms like the Multicast Broadcast Session Management Function (MB-SMF) and Policy Control Function (PCF).

\section{Related Work}
Several efforts have addressed reducing core traversal for multicast or group-based delivery in 5G.  Völk et al. ~\cite{volk2021emergency} proposed a Core–Edge Split EPC architecture for public safety networks, where essential core functions are deployed at the network edge in mobile base stations. This allows intra-cell communications to continue locally even when backhaul is unavailable, thereby avoiding core traversal and lowering latency for group voice/video traffic. The feasibility of such base-station–level routing aligns with our gNB-local breakout concept. In one of the earliest 5G Mobile Edge Computing (MEC) trials, Zhang et al. ~\cite{zhang2016mobile} quantified the latency benefits of offloading to a local MEC server compared to a remote core. Although MEC reduced one-way latency to about 17 ms, strict URLLC targets less than 5 ms remained out of reach, motivating further path shortening such as gNB-level forwarding. From the standards perspective, 3GPP TR 23.757 ~\cite{TR23757} examines Release 18 enhancements for MBS, including the concept of local MBS service where multicast delivery is restricted to a cell or tracking area. The study discusses placing multicast anchors closer to the RAN and even enabling RAN-based switching between unicast and multicast modes, reinforcing the idea that core traversal can be bypassed for local group traffic. Coll-Perales et al. ~\cite{coll-2022} analytically model end-to-end latency in V2X deployments with varying application server placement. Their results show centralized cloud deployments incur tens of milliseconds more delay than edge-hosted ones, confirming that moving processing and replication closer to the RAN yields substantial latency gains. This mirrors the $\approx$ 10 ms improvement we observe when replacing core-anchored multicast with gNB-local breakout. Complementing these approaches, Säily et al. ~\cite{saily20205g} developed a 5G NR RAN-controlled multicast architecture (5G-Xcast) with dynamic unicast/multicast switching and minimal 5GC impact. Their design demonstrates that multicast replication and group control can reside largely within the RAN.

Our work differs from these prior efforts in three key ways. First, unlike MEC approaches that still require an extra hop through a local core anchor, our method collapses the intra-cell data path entirely into the gNB while preserving 3GPP control-plane compliance. Second, while standards studies and RAN-centric frameworks consider multicast at or near the RAN, we explicitly target deterministic URLLC group control in industrial private 5G settings, with policy-driven eligibility conditions for local breakout. Third, we provide quantitative simulation evidence showing stable $\approx$ 10 ms latency reduction and sub-2 ms group delivery, directly linking these gains to the removal of the Backhaul/UPF/AF segment. This work makes the following contributions:
\begin{itemize}
  \item \textit{RAN–local multicast data path.} We design a gNB-local multicast breakout architecture that pivots eligible uplink flows $(s,f)$ to a downlink point-to-multipoint (PTM) bearer $B_g$ within the gNB that eliminates the core traversal for intra-cell group delivery while retaining core-anchored control.
  \item \textit{Standards alignment, security, and compliance.} We show how the work fits the 3GPP MBS/5GS architecture ~\cite{3gpp.23.247,3gpp.23.501} without redefining the core roles of authentication, ciphering, and integrity~\cite{3gpp.33.501}.
  \item \textit{Latency and scalability analysis.} We derive a latency decomposition and demonstrate that local intra-cellular breakout reduces latency $\approx$10\,ms. We validate the work using simulation that the average group latency remains 2\,ms within the group with a stable $\approx$10\,ms gap between local breakout and core-anchored delivery.
  \end{itemize}

\section{System Model and Problem Statement}
\label{sec:system}

\subsection{Network and Group Model}
\label{ssec:NW+groupModel}

We consider a private 5G deployment that covers an industrial cell served by a single gNB. Let $\mathcal{U}=\{1,\dots,N\}$ denote the set of user equipments (UEs) that serve the application such as robots, sensors, controllers, AGVs. attached to the cell. Devices subscribe to one or more \emph{industrial multicast groups} $\mathcal{G}\subseteq 2^{\mathcal{U}}$, where each $g\in\mathcal{G}$ is a subset of $\mathcal{U}$ authorized by policy. For any $g\in\mathcal{G}$, let $\mathcal{R}(g)\subseteq g$ be the receiver set and let $s\in g$ be a designated source that generates time-critical events or state updates to be distributed to $\mathcal{R}(g)$.

Traffic from $s$ to $g$ is carried out in a flow from uplink identified by $(s,f)$ with a per packet deadline $D$ and a reliability target $R$ (e.g. $D\le 5$, ms and $R\ge 99.999\%$). We measure end-to-end latency $L$ from MAC ingress at $s$ to MAC delivery at each $r\in\mathcal{R}(g)$; reliability is defined as $L$;
\begin{equation}
\Pr\big[\,L \le D\,\big] \;\ge\; R,
\end{equation}

\subsection{Local Multicast Forwarding at the gNB}
\label{ssec:localForwarding}

The gNB hosts a \emph{local user-plane function} (local UPF) that maintains a forwarding table
\[
\mathrm{FT}:\ (s,f)\ \mapsto\ (g, B_g),
\]
where $B_g$ denotes a configured point-to-multipoint (PTM) downlink bearer serving $\mathcal{R}(g)$. Upon receiving an eligible uplink PDU for $(s,f)$, the gNB \emph{pivots} the payload to $B_g$ and schedules a PTM transmission to $\mathcal{R}(g)$ within the next available PDSCH slot. The control-plane functions (admission, authorization, group membership, and policy) remain anchored in the 5GS core. Only the \emph{data} path for intra-cell dissemination is locally broken out in the gNB.
Eligibility for local breakout is governed by policy. When eligibility fails, the gNB is returned to the standard core-anchored path without interruption of service.

\subsection{Latency Decomposition}
For intra-cell group delivery, the end-to-end latency under core-anchored multicast can be decomposed as

\begin{align}
L_{\mathrm{CA}} =\; & T_{\mathrm{rqt}} + T_{\mathrm{UL}} 
+ T_{\mathrm{gNB\_proc}} \nonumber \\
& + T_{\mathrm{BH/UPF/AF}} 
+ T_{\mathrm{DL\_schd}} + T_{\mathrm{DL}}
\end{align}

where $T_{\mathrm{BH/UPF/AF}}$ aggregates backhaul, UPF, application processing, and associated queueing. Under local forwarding,
\begin{equation}
L_{\mathrm{LB}} \;=\; T_{\mathrm{rqt}}+T_{\mathrm{UL}} + T_{\mathrm{gNB\_proc}} + T_{\mathrm{DL\_schd}} + T_{\mathrm{DL}}
\end{equation}
Payloads do not traverse the core user-plane. The design thus aims to eliminate $T_{\mathrm{BH/UPF/AF}}$ minimizing the latency.


\noindent
\textbf{Assumptions.} (i) Single-cell analysis: inter-cell groups revert to core-anchored MBS or use inter-gNB coordination. (ii) Control-plane policies are distributed to the gNB; the local MB-UPF does not weaken authentication, ciphering, or integrity protection. (iii) Mobility events trigger a seamless transition between the local and core paths.

\noindent
\textbf{Problem Statement.} Given $\mathcal{U}$, $\mathcal{G}$, deadlines $D$, reliability goals $R$, and gNB’s scheduling resources, design a gNB-local multicast forwarding policy $\pi$ that (a) pivots eligible uplink flows $(s,f)$ to the downlink PTM bearer $B_g$ and (b) schedules transmissions to minimize latency while meeting reliability, and security. The hypothesis is that $L_{\mathrm{local}} \ll L_{\mathrm{core}}$ for intra-cell groups that yields order-of-magnitude improvements in the presence of nontrivial $T_{\mathrm{BH/UPF/AF}}$.

\section{Proposed gNB Local Multicast Forwarding}
\label{sec:local-multicast}

\subsection{High-Level Architecture}
\label{ssec:highLevelArchitecture}

The control-plane flow is considered as defined by 3GPP. The MB-SMF, 5G core function, is used for session admission and group or policy dissemination and AMF/PCF cooperation. Only relocating the \emph{intra-cell} user-plane replication from the core MB-UPF to the gNB \cite{3gpp.23.247,3gpp.23.501,3gpp.23.502}. Concretely, the gNB hosts a MB-UPF that maintains a local forwarding map for multicast traffic.
\[
\mathrm{FT}:\ (s,f)\ \mapsto\ (g, B_g)
\]
where $(s,f)$ identifies an authorized uplink traffic flow from source $s$, $g$ to the multicast group, and $B_g$ is a configured point-to-multipoint (PTM) downlink bearer serving $\mathcal{R}(g)$ on NR as per 3GPP TS 38.300~\cite{3gpp.38.300}. For an eligible uplink PDU, the local MB-UPF on gNB pivots the payload to $B_g$ at PDCP/RLC and schedules PTM PDSCH within the next feasible slot, thus avoiding core traversal of the \emph{data} path while preserving core anchored control and policy.

Group membership and \emph{local-breakout} permissions are distributed to the gNB through MB-SMF/AMF under PCF policy \cite{3gpp.23.247,3gpp.23.501}. The local MB-UPF considers a flow pivoting $(s,f)$ eligible if: (i) all intended receivers are currently attached to the serving cell (ii) PRB reservations admit PTM transmission. If any condition fails, the gNB transparently reverts to core-anchored MBS without service interruption.

\subsection{Algorithm description and fallback}
\label{ssec;algorithms}

The gNB maintains a forwarding table keyed by UE and traffic identifiers to enable fast local multicast. When a packet arrives, it checks if a forwarding entry exists and if local breakout is permitted. If either check fails, the packet is passed to the core and served via the standard MB‑UPF path. Otherwise, the associated multicast group and bearer are retrieved, and the payload is queued in the appropriate PDCP. Select PTM on the next available PDSCH transmission slot. This approach delivers ultra‑low‑latency multicast within a cell while reverting to the core when UEs move across cells or policies change.

Alogrithm \ref{alg:lma} provides a high-level view of this local multicast forwarding process at the gNB. It shows how the forwarding table (FT) and policy checks determine whether an uplink PDU is sent directly to the core or pivoted into the appropriate downlink multicast bearer for PTM transmission.


\begin{algorithm}[tbh]
\caption{gNB Local Multicast Breakout}
\label{alg:lma}
\begin{algorithmic}[1]
\small
\Require UL~PDU $(UE\_id,\, flow\_id,\, payload)$
\Statex \textbf{State:} FT (forwarding table), Policies $\Pi$, Group $G$, Multicast Bearer $B_G$
\If{$(UE\_id,\, flow\_id) \notin FT$}
  \State \textbf{SendToCore}(); \Return
\EndIf
\If{$\neg\, \Pi.\textsc{LocalBreakoutAllowed}(UE\_id,\, flow\_id)$}
  \State \textbf{SendToCore}(); \Return
\EndIf
\State $G \gets FT[UE\_id,\, flow\_id].group$
\State $B_G \gets FT[UE\_id,\, flow\_id].bearer$
\State \textbf{EnqueueToPDCP}$(payload,\, B_G,\, \text{QoS\_marking})$
\State \textbf{SchedulePTM\_PDSCH}(\text{NextEligibleSlot\-AlgnToWnd})
\State \textbf{CollectLimitedNAKs}()
\If{\textbf{AnyNAKs}()}
  \State \textbf{SelectiveUnicastRepair}()
\EndIf
\end{algorithmic}
\label{alg:lma}
\end{algorithm}

\section{Method and Discussion}
\label{sec:method}

We model a single private 5G cell with one gNB and up to $N{=}150$ UEs uniformly distributed within a radius 100\,m. The gNB is at $(0,0,30)$ with 4 antennas and each UE has 2 antennas. NR uses 30\,kHz SCS (\emph{slot} $=0.5$, ms), $100$, RB over 100\,MHz in 3.5\,GHz, up to 2 MIMO layers and 64QAM in UL/DL. Each UE generates on/off traffic using \texttt{networkTrafficOnOff} with \texttt{OnTime}$=10$\,ms, \texttt{OffTime}$=90$\,ms, \texttt{DataRate}$=1$\,Mbps, and packet size $\approx 1002$, bits. Inter-arrival times and payloads are sampled per slot; packets are bit-padded to satisfy NR layer mapping constraints. A CDL-D (\texttt{nrCDLChannel}) model is used with carrier-consistent sampling, maximum Doppler 10\,Hz, and per-link seeds for repeatability. Decoders are invoked with zero noise variance to isolate path latency from PHY (physical layer) errors.
The uplink uses unicast PUSCH (Physical Uplink Shared Channel) from each UE to the gNB whereas downlink uses PTM PDSCH multicast to all receivers. PRB sets span the entire bandwidth for clarity of comparison. The MATLAB simulation implementation is available on: ~\nolinkurl{https://github.com/rajendra1124/LocalBreakout}. 

\noindent
\textbf{Scenarios:} 
\emph{Core-anchored MBS:} gNB$\!\to$MB-UPF/AF$\!\to$gNB adds a fixed user-plane delay of 5-12\,ms per multicast packet~\cite{latency_tech}. The variation in delay is based on the network design and configuration. \\  
\emph{Local breakout:} gNB pivots eligible UL payloads to the DL PTM bearer locally, and radio parameters are identical across both paths.

\subsection{Latency Analysis}
We decompose end-to-end latency $L$ for intra-cell group delivery:
\begin{align}
L &= T_{\mathrm{UL\_tx}} + T_{\mathrm{gNB\_proc}} + T_{\mathrm{path}} + T_{\mathrm{DL\_schd}} + T_{\mathrm{DL\_tx}}
\end{align}

\begin{figure}[t]
~\label{sec:lat_comp}
  \centering
  \includegraphics[width=0.92\linewidth]{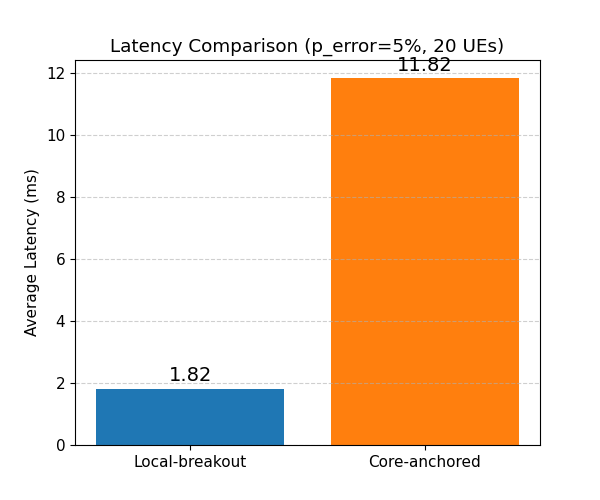}
  \caption{Average latency of Core-anchored Vs gNB local breakout forwarding for intra-cell groups.}
  \label{fig:arch2}
\end{figure}

From Table \ref{tab:lmb} and according to (4), the end-to-end latency decomposes into four parts for intra-cell group delivery under 30 kHz SCS. Figure \ref{fig:arch2} contrasts the two delivery paths and highlights that removing the Backhaul / UPF / AF segment with gNB-local forwarding yields a stable $\approx$ 10 ms reduction in end-to-end latency across group sizes, which matches the decomposition in (4). UL grant + tx (0.25–1.0 ms) covers access and PHY transmission with request-based UL and queuing.  A gNB processing (1.0–2.0 ms) accounts for PDCP/RLC handling, group mapping, and local breakout and PTM scheduling. The Backhaul + UPF + AF segment 5.0–10.0 ms of transport and core processing exists only for core-anchored MBS and disappears with local breakout because payloads never leave the gNB user-plane. DL scheduling + tx (0.25–1.0 ms) reflects the wait to the next PTM opportunity and PDSCH time which get shortened by mini-slot alignment and reserved PRBs. Summing these yields 6.5 to 11.5 ms for core-anchored MBS versus 1.5 to 4.0 ms for local breakout. The ranges are primarily driven by numerology, scheduler protection, and any selective repair or duplication mechanism. 

\begin{figure}[t]
~\label{sec:lat_comp}
  \centering
  \includegraphics[width=1\linewidth]{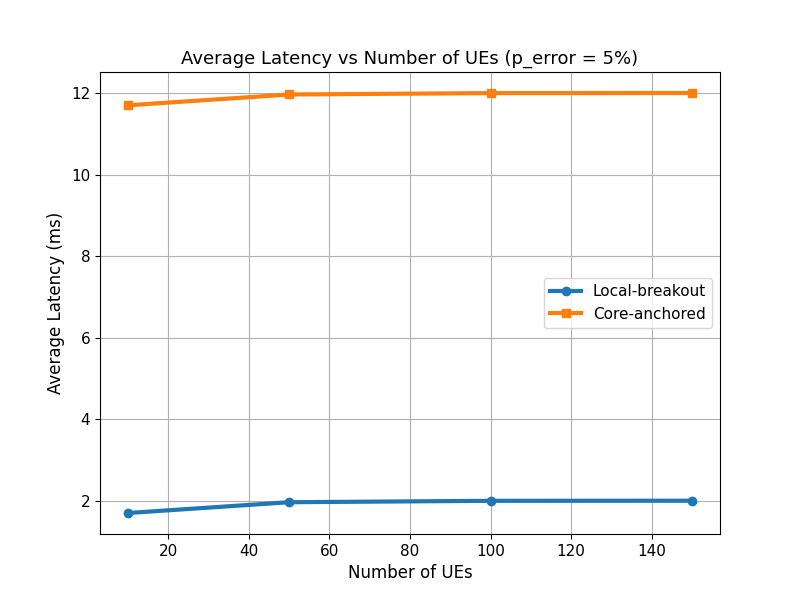}
  \caption{Latency comparison of Core-anchored Vs local breakout when number of UEs changes.}
  \label{fig:arch5}
\end{figure}

Figure \ref{fig:arch5} shows that the local breakout architecture consistently achieves a group latency of sub 2 ms between 10 to 150 receivers, while the core-anchored architecture remains around 12 ms on average well above a URLLC deadline of 5 ms. The nearly flat trend with group size reflects the efficiency of PTM (one DL transmission for all receivers). The constant gap between the curves of 10 ms is the removed core user-plane segment that the local breakout method eliminates. Overall, the figure substantiates two claims: (i) intra-cell local breakout meets tight industrial deadlines with comfortable margin, and (ii) multicast scales with group size without degrading average latency. 

For private 5G in Industry~4.0/5.0, the local-breakout method provides a practical path to deterministic group dissemination without architectural upheaval. The pivoting of data from the user-plane moves to the gNB, while the control-plane remains as per the 3GPP MBS architecture~\cite{3gpp.23.247}. 


\begin{table}[t]
\centering
\caption{Indicative latency (intra-cell group delivery).}
\label{tab:budget}
\begin{tabular}{lcc}
\toprule
Component & Core-anchored MBS & Local breakout \\
\midrule
UL grant + tx & 0.25--1.0 ms & 0.25--1.0 ms \\
gNB processing & 1.0--2.0 ms & 1.0--2.0 ms \\
Backhaul + UPF + AF & 5.0--10.0 ms & \textbf{0} \\
DL scheduling + tx & 0.25--1.0 ms & 0.25--1.0 ms \\
\midrule
\textbf{Total} & \textbf{6.5--11.5 ms} & \textbf{1.5--4 ms} \\
\bottomrule
\end{tabular}
\label{tab:lmb}
\end{table}

Our design preserves 3GPP security and compliance guarantees while adding the local UPF function collocated at the gNB to connect industries 4.0/5.0 scenarios into next-generation networks. The gNB’s local breakout changes the location of replication from the core to the gNB, while leaving the security anchors unchanged. The authorization, group membership, and policy stay anchored in the core. So authentication, ciphering, and integrity protection remain exactly as specified in 3GPP TS 33.501/23.501 ~\cite{3gpp.33.501,3gpp.23.501}. 

We acknowledge limits and threats to validity, as the results are from a single-cell setting with controlled traffic and simplified radio/error models. The multi-cell operation, inter-cell groups, handover transients, and correlated interference could change the latency margin. Hardware contention or misconfiguration on the gNB could affect isolation. Future work should therefore evaluate multi-cell deployments with realistic mobility and channel dynamics.

\section{Conclusions and Future Works}
\label{sec:conclusions}

We presented a gNB-local breakout architecture for industrial URLLC in private 5G that pivots eligible uplink traffic at the gNB to a downlink PTM bearer, while leaving authorization, membership, and policy anchored in the core. This local forwarding, commonly referred to al 'local breakout', collapses the intra-cell data path by removing the Backhaul/UPF/AF segment. Hence the end-to-end group latency reduced to roughly sub 2 ms averages in our settings. There is also a stable $\approx$10 ms gap to the core-anchored delivery across loads and group sizes. Crucially, the method preserves 3GPP security and compliance: authentication, ciphering, and integrity. This method is practical for coordinated robotics, AGVs, machine vision, and human–robot collaboration in Industry 4.0 / 5.0 that require low single-digit latency. Although our results are from a single-cell model with controlled traffic, results indicates RAN-local data forwarding is a simple, effective for URLLC grade group dissemination. 

Our ongoing work focuses on explicit HARQ and Negative Acknowledgment feedback and selective unicast repair. We are taking it to the next level by extending to multi-cell setups, ensuring smooth handover and gNB coordination with a fallback to default core routing for reliability in busy Industry 5.0 settings. To nail down that five-nines reliability, we’re going to use probabilistic models and smart tools to predict and dodge packet failures, even in tough network conditions, ensuring steady performance for small data packets.
Also, prototyping the scheduler on a real gNB/O-RAN platform to validate latency, observability, and compliance under realistic interference and load.


\bibliographystyle{IEEEtran}
\bibliography{sample}
\end{document}